\documentclass[aps,twocolumn,showpacs,preprintnumbers,amsmath,amssymb,floatfix]{revtex4}

\usepackage{graphicx}
\usepackage{dcolumn}
\usepackage{bm}
\usepackage[german,american]{babel}

\begin{document}


\title{Dynamics of liquid silica as explained by properties of the potential energy landscape}

\author{A. Saksaengwijit}
\author{A. Heuer}%

\affiliation{
Westf\"{a}lische Wilhelms-Universit\"{a}t M\"{u}nster, Institut f\"{u}r Physikalische Chemie
Corrensstr. 30, 48149 M\"{u}nster, Germany }

\date{\today}

\begin{abstract}
The dynamics of silica displays an Arrhenius temperature
dependence, classifying silica as a strong glass-former. Using
recently developed concepts to analyse the potential energy
landscape one can get a far-reaching understanding of the
long-range transport of silica. It can be expressed in terms of
properties of the thermodynamics as well as local relaxation
processes, thereby extending the phenomenological standard picture
of a strong glass-former. The local relaxation processes are
characterized by complex correlated sequences of bond breaking and
reformation processes.
\end{abstract}

\pacs{64.70.Pf, 65.40.Gr, 66.20.+d} 
\maketitle

Silica is a prototypical and technologically relevant glass-former,
displaying a variety of remarkable physical properties like
thermodynamic anomalies\cite{Saika:2001,Shell:2002,Saika:2004}. In
contrast to most other glass-formers the temperature dependence of
its transport properties like the oxygen self-diffusion constant
$D(T)$ display a simple Arrhenius behavior with an activation energy
$V_{diffusion} = $ 4.7 eV \cite{Mikkelsen:1984} and is thus a strong
glass-former \cite{Angell:1995,Ruocco:2004}.  This suggests that the
transport can be described as a successive breaking and reformation
processes of Si-O bonds with an activation energy close to
$V_{diffusion}$ \cite{Ediger:2000,Debenedetti:2001}.

To scrutinise this simple picture and thus to obtain a microscopic
picture of the dynamics of silica  we employ the framework of the
potential energy landscape (PEL), defined in the high-dimensional
configuration space \cite{Debenedetti:2001}. At low temperatures the
properties of silica and other glass forming systems are mainly
characterized by the properties of the local potential energy minima
of the PEL (denoted inherent structures, IS)
\cite{Stillinger1:1982,Debenedetti:2001}. The thermodynamics of the
system is mainly governed by the energy distribution of the number
of IS. Introducing the configurational entropy $S_c(T)$ as a measure
for the number of accessible IS at a given temperature, there is an
empirical connection of $S_c(T)$ to the dynamics $D(T) \propto
\exp(-A/TS_c(T))$ (Adam-Gibbs relation \cite{Adam-Gibbs:1965}) with
some fitting parameter $A$ \cite{Sastry:2001,Saika:2001,
LaNave:2002}. Its theoretical foundation, however, is under
debate\cite{Wolynes:2001,Garrahan:2003} and no direct interpretation
of $A$ is available, yet. In any event, one would expect that also
the topology of the PEL should be of utmost importance for
understanding the dynamics \cite{Debenedetti:2002}.

Following the ideas of Stillinger and Weber \cite{Stillinger1:1982}
one can map a trajectory, obtained via a molecular dynamics
simulation, to a sequence of IS via frequent minimization of the
potential energy. One example is shown in Fig.1. This highlights how
the system is exploring the PEL. One can group together IS to
metabasins (MB) such that the dynamics between MB does not contain
any correlated backward-forward processes
\cite{Debenedetti:2001,Doliwa_Eacc:2003,Denny:2003}. Each MB is
characterized by a waiting time $\tau$ and an energy $e$ (defined as
the lowest IS energy in this MB). The technical details of this
approach, which have been developed in the context of studying a
simple glass-forming model, i.e. the binary mixture Lennard-Jones
system (BMLJ), can be found in \cite{Doliwa_Eacc:2003}.

\begin{figure}
\includegraphics[width=7cm]{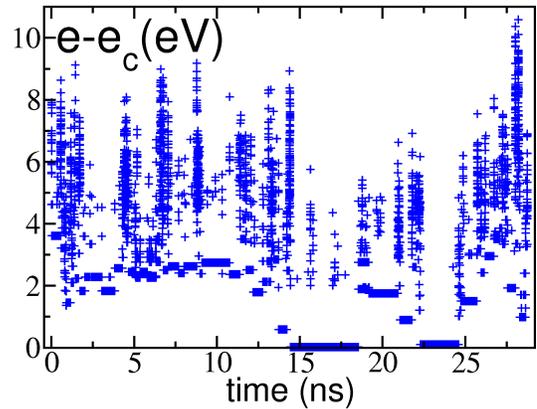}
\caption{\label{fig1} Time dependence of the IS energy e during a
molecular dynamics simulation at T = 3000 K. The fountain-like
objects correspond to time periods during which the system is moving
very fast in configuration space. $e_c$ is an estimate for the
low-energy cutoff of the PEL.}
\end{figure}

In this work we apply these techniques to silica. Combining
information about the energy distribution of IS and the local
relaxation processes, reflecting the local topology of the PEL, we
obtain a far-reaching understanding of its dynamics. From this we
can identify the reasons why silica is a strong system and obtain
a quantitative understanding of observables like the resulting
activation energy $V_{diffusion}$. Further information is obtained
from an appropriate comparison with the real-space behavior of
silica. The underlying picture, emerging from these results,
extends substantially the rationalization of the strong behavior
of silica, sketched above.

Information about the BKS potential \cite{BKS:1990}, used to model
silica, as well as further simulation details can be found in Ref.
\cite{Heuer:2004}. For an optimum analysis in terms of the PEL the
system size should be as small as possible without showing
significant finite size effects \cite{Doliwa_Eacc:2003}.  It has
been shown that already for system sizes $N \approx 10^2$ finite
size effects concerning the configurational entropy
\cite{Heuer:2004}, the properties of tunneling systems
\cite{Jens:2005}, the temperature dependence of the oxygen diffusion
\cite{Heuer:2004} as well as the nature of the relaxation processes
in BKS silica (checked, e.g., via the degree of non-exponentiality
in the incoherent scattering function)
\cite{Heuer:2004,Martin-Samos2:2005}are small in the accessible
range of temperatures. Here we choose $N=99$. Properties of larger
systems can be then predicted from statistical arguments
\cite{Wales:2003b}. Recent studies have shown that the distribution
of configurational energies has a low-energy cutoff around some
energy $e_c$ \cite{Heuer:2004} with a finite configurational entropy
\cite{Saika:2004}. It results from the network constraints in
defect-free configurations \cite{Heuer:2004}. It will turn out to be
one key feature for the understanding of the dynamics.

In analogy to previous results for the BMLJ system
\cite{Doliwa_Hop:2003} the oxygen diffusion constant $D(T)$ is
proportional to the inverse of the average MB waiting time $\langle
\tau(T) \rangle$; see Fig.2. Thus, a local quantity like $\langle
\tau(T) \rangle$ fully determines the temperature dependence of
diffusion, i.e. $d \ln D(T) / d \beta$ . The low-temperature
activation energy $V_{diffusion} = 4.84$ eV is very close to the
experimentally observed value of 4.7 eV \cite{Mikkelsen:1984}.
Around 3500 K one observes the crossover from the high-temperature
non-Arrhenius to the low-temperature Arrhenius-regime
\cite{Horbach:1999,Saika:2001,LaNave:2002,Saika:2004}.

\begin{figure}
\includegraphics[width=6.4cm]{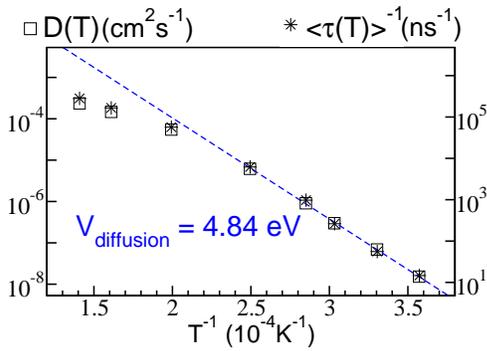}
\caption{\label{fig2} Temperature dependence of the oxygen diffusion
constant $D(T)$ (macroscopic transport) and the inverse average
waiting time $\langle \tau(T) \rangle^{-1}$ (microscopic
relaxation). The low-temperature activation energy is $V_{diffusion}
= 4.84$ eV. Around 3500 K the so-called fragile-to-strong crossover
is observed \cite{Horbach:1999,Saika:2001,LaNave:2002,Saika:2004}. }
\end{figure}

One may suspect that the MB waiting times are strongly energy
dependent because high-energy configurations have a larger number of
defects \cite{Heuer:2004} which can more easily relax
\cite{Jens:2005}. For a quantification we determine the average MB
waiting time in dependence of temperature {\it and} energy, denoted
$\langle \tau(e,T) \rangle$ from analysis of several independent
equilibrium runs; see Fig.3. Interestingly, for all $e$ one finds a
simple Arrhenius-behavior, characterized by an effective activation
energy $V_{MB}(e)$ and a prefactor $\tau_0(e)$. Thus, the
Arrhenius-behavior is not only present for simple atomic systems
like the BMLJ system \cite{Doliwa_Eacc:2003} but also for silica.
One observes a crossover from the high-energy regime with $V_{MB}(e)
\approx V_0$ and $\tau_0(e) = \tau_{micro}$ to a low-energy regime
with a strong energy-dependence of both functions. The resulting
broad distribution of waiting times implies that energy is likely
the most dominating factor for understanding the occurrence of
dynamic heterogeneities in silica and other glass-forming systems
\cite{Vogel:2004}.

\begin{figure}
\includegraphics[width=7.4cm]{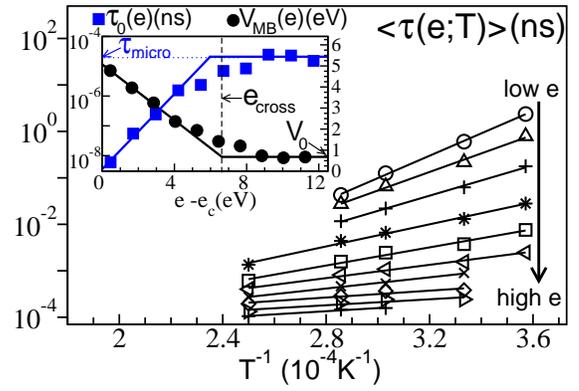}
 \caption{\label{fig3}  Arrhenius plot of the average MB waiting time
$\langle \tau(e,T) \rangle $ in dependence of energy (the chosen
energy bins can be read off from the inset). $\langle \tau(e,T)
\rangle$ can be characterized by an effective activation energy
$V_{MB}(e)$ and a prefactor $\tau_0(e)$, shown in the inset. There
exists a dynamic crossover energy $e_{cross}$ such that for $e
> e_{cross}$, $\tau_0(e)=\tau_{micro} \approx 20$ fs and $V_{MB}(e) = V_0 \approx 0.8$ eV are
roughly constant. The lines are guides to the eyes.}
\end{figure}

For relating the thermodynamics and the dynamics we introduce
$p(e,T)$ as the Boltzmann probability to be in a MB of energy $e$.
It turns out that in the relevant energy and temperature range
$p(e,T)$ is virtually identical for MB and IS \cite{Heuer:2004} and
thus contains all the information about the configurational entropy.
Using the quantities, introduced so far, we can write down the
formal relation \cite{Doliwa_Eacc:2003}
\begin{eqnarray}
D(T) \propto \langle \tau(T) \rangle^{-1} & = &  \int \, de
p(e,T)/\langle \tau(e,T) \rangle \nonumber \\ &= &  \int \, de
p(e,T) \frac{\exp(-\beta V_{MB}(e))}{\tau_0(e)} .
\end{eqnarray}
Its physical relevance is far-reaching because it relates the
thermodynamics (via $p(e,T)$) and the local dynamics (via $\langle
\tau(e,T) \rangle$) to the long range diffusion. It follows that for
very low temperatures for which $p(e,T)$ has only contributions for
$e \approx e_c$ the dominant contributions to the average waiting
time originate from configurations with energies close to $e_c$ via
$\langle \tau(e \approx e_c,T) \rangle$. Then the local Arrhenius
behavior $\langle \tau(e \approx e_c,T) \rangle \propto \exp(\beta
V_{MB}(e\approx e_c))$ translates into a macroscopic Arrhenius
behavior. Indeed, the macroscopic activation energy $V_{diffusion}$
is close to $V_{MB}(e \approx e_c)$; see Fig.4.

What determines the value of the crossover temperature of 3500 K? At
this temperature $p(e,T)$ is peaked around $e_c + 4$ eV and the
low-energy wing of $p(e,T)$ just starts to be influenced by the
low-energy cutoff \cite{Heuer:2004}. Thus, on first view the above
arguments to rationalize Arrhenius behavior should only apply for
lower temperatures. However, due to the additional weighting of
$\exp(\beta V_{MB}(e))$ by $1/\tau_0(e)$ in Eq.1, which for $e
\approx e_c$ is more than four orders of magnitude larger than
$1/\tau_{micro}$ ($\tau_0(e \approx e_c) \approx 5 \cdot 10^{-3}$
fs), the influence of the low-energy states in the integral is
significantly enhanced, thus giving rise to the actually observed
crossover. Is there a direct relation to the mode-coupling
temperature $T_c \approx 3300 K$ \cite{Horbach:1999}? Qualitatively,
$T_c$ is related to the beginning dominance of activated processes
rather than to the presence of a low-energy cutoff of the PEL and
thus seems to have a different physical origin.

In the next step we elucidate the {\it microscopic} origin of the
escape properties from configurations close to $e_c$. We present
detailed results for one specific configuration (denoted IS$_0$). In
a first step we determine its average waiting time via repeated runs
starting at IS$_0$ with different initial velocities and different
temperatures. These runs typically involve many unsuccessful escape
attempts. We define the waiting time as the time when IS$_0$ is left
for the last time. We find a simple Arrhenius behavior with an
effective activation energy of $V_{escape} = 4.77$ eV (Fig.4a).

\begin{figure}
\center{
\includegraphics[width=8.8cm]{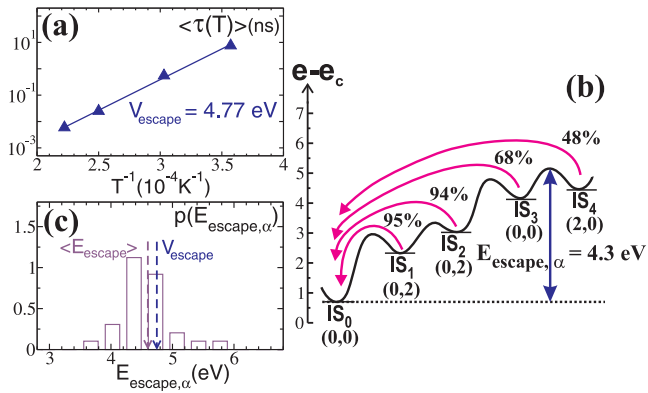}
\caption{\label{fig4} Detailed analysis of the escape properties
from one defect-free low-energy structure IS$_0$. a) Temperature
dependence of the average waiting time in IS$_0$. b) One specific
escape path starting from IS$_0$. This representation reflects the
sequence of inherent structure energies as well as saddle energies.
The results for $p_{back}$ are given above the arrows. The pairs of
numbers at the bottom reflect the numbers of silicon and oxygen
defects, respectively.  The resulting effective barrier height
$E_{escape,\alpha}$ is indicated. c) Distribution of
$E_{escape,\alpha}$ and its average value $\langle E_{escape}
\rangle = 4.6$ eV ($\approx V_{escape}$) for 30 different escape
paths from IS$_0$. } }
\end{figure}

This activation energy incorporates all the complexity of the local
PEL around this configuration. In particular, the configuration may
relax along different paths $\alpha$, which all contribute to the
total relaxation rate, i.e. $\Gamma = \sum \Gamma_\alpha$. If
$p_\alpha$ is the probability to escape via path $\alpha$ and if
path $\alpha$ is characterized by an effective barrier height
$E_{escape,\alpha}$ (see below for its definition) one can write
$\langle E_{escape}\rangle \equiv \sum p_\alpha E_{escape,\alpha}$
\cite{Doliwa_Eacc:2003}, which is the average value of
$E_{escape,\alpha}$ for independent runs starting from IS$_0$. A
typical escape path is shown in Fig.4b. One can see the sequence of
IS after IS$_0$ is visited for the last time. For every IS$_i$ a
value, denoted $p_{back}$, has been obtained from counting for a set
of independent simulations with starting structure IS$_i$, whether
or not the system returns to IS$_0$. Qualitatively, $p_{back} < 0.5$
indicates that the system will (on average) escape from the
catchment region of IS$_0$. This limit typically involves the
entropic effect that due to the large number of transition options
in the high-dimensional PEL there is no need to follow the path back
to IS$_0$. As shown in Ref.\cite{Doliwa_Eacc:2003} for a given
escape path a good estimate of $E_{escape,\alpha}$ is first to
locate the first IS with $p_{back} < 0.5$ and then to identify
$E_{escape,\alpha}$ as the highest energy along the reaction
coordinate up to this IS. Application to the escape path in Fig.4b
yields $E_{escape,\alpha} = 4.3$ eV. Four transitions are required
until for the first time $p_{back} < 0.5$.

From the distribution of $E_{escape,\alpha}$ for 30 different escape
paths from IS$_0$ one obtains $\langle E_{escape} \rangle = 4.6$ eV
which is close to the value of $V_{escape} = 4.77$ eV ; see Fig.4c.
Thus,  it is indeed possible to quantitatively relate the effective
activation energy to the local properties of the PEL. Repeating this
analysis for two different low-energy IS we get a similar agreement.
More generally this implies that $V_{MB}(e \approx e_c)$, on the one
hand, can be quantitatively related to specific local barriers and,
on the other hand, to the activation energy of macroscopic diffusion
$V_{diffusion}$. This establishes a direct link between the
microscopic and macroscopic behavior of silica.

Having identified a complete escape process (relevant for
$V_{diffusion}$) via the $p_{back}$-criterion we are in a position
also to analyze its {\it real-space} characteristics. First we note
that in the vast majority of cases the sequence of IS-transitions
during the escape is correlated. This is reflected by the fact that
at least one atom, involved in a bond-breaking or reformation
process during a specific IS-transition ($i\ge 1$) IS$_i \rightarrow
$IS$_{i+1} $ was involved during a previous IS-transition. When
comparing IS$_4$ with IS$_0$ in total 4 Si-O bonds have been broken
and 5 silicon atoms have changed their oxygen neighbors. These
values are close to the average behavior after analyzing the escape
from all three three initial IS (4.6 and 4.4, resp.). This implies
significant correlated bond-breaking and reformation processes. In
particular, $V_{diffusion}$ cannot be related to the breaking of a
single Si-O bond. Other researchers have rationalized the value of
$V_{diffusion}$ by the sum of half of the mean formation energy of
an oxygen Frenkel pair and a migration barrier
\cite{Martin-Samos2:2005}. On a qualitative level something similar
is observed here because in the first step a defect is created and
afterwards the defect is transferred until $p_{back} < 0.5$. A
closer look, however, reveals that the behavior in BKS silica is
more complex as reflected, e.g., by the fact that in the example of
Fig.4 in between also configurations with no defects occur (IS$_3$).

The additional contribution of the saddle between IS$_3$ and IS$_4$
of approx. 0.8 eV to the final value of $\langle E_{escape} \rangle
$ is small. This also holds in general (approx. 1.0 eV) which,
interestingly, is close to $V_0$. Thus one may conclude that there
are two distinct contributions to the activation energy $V_{MB}(e)$:
(1) $V_{MB}(e) - V_0$ as the contribution reflecting the topology of
the PEL, related to differences between IS energies, and (2) $V_0$
as the additional contribution of the final saddle. Around $e
\approx e_c$ the first contribution is dominating.

Why is silica a strong glass-former?  The standard scenario,
sketched in the introductory paragraph, would imply  $\langle \tau
(e,T) \rangle \approx \tau_{micro} \cdot \exp(\beta V_{diffusion})$
\cite{Ediger:2000,Debenedetti:2001}. This reflects the presence of
one typical relaxation process which holds throughout the entire PEL
\cite{Debenedetti:2001}. In contrast, our simulations have revealed
a strong energy dependence of $V_{MB}(e)$ and $\tau_0(e)$  together
with complex successive bond-breaking and reformation processes.
Rather we can identify two underlying reasons for the classification
of silica as a strong glass-former: (1) the presence of the cutoff
in the PEL of silica as a consequence of its network structure, (2)
the Arrhenius temperature dependence of $\langle \tau(e\approx
e_c,T)\rangle$ together with a large attempt rate $1/\tau_0(e_c)$.
How to rationalize property (2)? As can be seen from Fig.4c the
distribution of effective barrier heights $E_{escape,\alpha}$ is
very narrow. For the example of IS$_0$ this implies an Arrhenius
temperature dependence in agreement with the observation; see
Fig.4a. This behavior was also observed for the escape from the
other low-energy IS, analyzed along the same lines. Thus, it seems
to be a general feature that starting from a low-energy (and
typically defect-free) configuration around $e_c$ the system first
has to acquire an energy of approx. $e \approx e_c + V_{MB}(e_c)$
until the escape is complete (ending in an IS with $e \approx e_c +
V_{MB}(e_c) -V_0$). More pictorially one may state that low-energy
IS ($e \approx e_c$) form the bottom of crater-like objects in the
PEL and the system has no escape option apart from climbing up the
whole crater until $e \approx e_c + V_{MB}(e_c)$. Thus, beyond the
presence of the low-energy bound of the PEL the strong behavior of
silica is also related to this crater-like structure of the PEL.

The previous thermodynamic analysis has revealed that for this
system the number of IS with $e_c + V_{MB}(e_c) -V_0 \approx e_c +
4$ eV is approx. $10^5$ times larger than the corresponding number
at $e \approx e_c$ \cite{Heuer:2004}. This observation suggests that
the number of possible transitions from $e \approx e_c$ to
configurations with energies around $e_c + 4$ eV is also
exponentially large, thereby rationalizing the dramatic increase of
$1/\tau_0(e_c)$ as compared to $1/\tau_{micro}$. It has been already
explicitly shown in previous model calculations that the prefactor
for relaxation processes in model landscapes scales with the number
of accessible states for which for the first time $p_{back}$ is
smaller than approx. 50\% \cite{aimorn:2003}. Then $1/\tau_0(e_c)$
can be much larger than microscopic jump rates. Interestingly, in
the relevant energy regime of the BMLJ system the number of IS
increases much weaker with increasing energy and correspondingly
there is hardly any energy-dependence of $\tau_0(e)$
\cite{Doliwa_Eacc:2003}. This further supports our hypothesis about
the origin of the large energy-dependence of $\tau_0(e)$.

It has been speculated that different network-forming systems (e.g.
silica and water) have similar
properties\cite{Angell:1999,Scala:2000,Stanley:2003}. Indeed,
indications have been found recently that also the amorphous states
of water possesses a low-energy cutoff, which will influence the
thermodynamics and dynamics at low temperatures similarly to
silica\cite{Poole:2005}. Therefore exploration of the properties of
silica may be of major importance also for an improved understanding
of water and other network formers. The possible universality of
this class of systems has recently lead to the formulation of an
abstract model which is aimed to reflect the basic physics of all
amorphous network formers \cite{Moreno:2005}.

In summary, using the PEL-framework we have identified relevant
underlying mechanisms for the dynamics of silica: (1) The elementary
relaxation processes are not simple bond-breaking processes with the
final activation energy but rather form correlated sequences of many
bond-breaking processes with a strong dependence on the initial
energy. (2) The {\it simultaneous} presence of the low-energy cutoff
of the PEL as well as the narrow distribution of escape barriers
from MB give rise to the resulting strong behavior. (3) Dramatic
entropic effects are present for the escape from low-energy states,
showing that the dynamics is much more complex than reflected by the
resulting simple Arrhenius behavior. (4) The occurrence of the
crossover temperature around 3500 K can be quantitatively understood
by the presence of the low-energy cutoff {\it and} these entropic
effects. (5) There exists a crossover energy which separates the
high-energy liquid-like behavior and the low-energy activated
behavior.

We~gratefully acknowledge helpful discussions~with
  R.~D.~Banhatti, B.~Doliwa, D.R.~Reichman, O.~Rubner and M.~Vogel. The work was supported in part by the DFG~via SFB~458
  and by the NRW Graduate School of Chemistry.


\end{document}